\documentclass[%
 aip,
 amsmath,amssymb,floatfix,
preprint,%
]{revtex4-1}

\usepackage{graphicx}
\usepackage{dcolumn}
\usepackage{bm}

\usepackage[utf8]{inputenc}
\usepackage[T1]{fontenc}
\usepackage{mathptmx}
\usepackage{etoolbox}
\usepackage{subfigure}
\usepackage{hyperref}
\usepackage{siunitx}
\usepackage[final]{changes}

\begin{document}


\title{Direct numerical simulation of flapping flags in grid-induced turbulence}



\author{Stefano Olivieri}
\affiliation{
 Complex Fluids and Flows Unit, Okinawa Institute of Science and Technology Graduate University, 1919-1 Tancha, Onna-son, Okinawa 904-0495, Japan
}
\email[Corresponding author: ]{stefano.olivieri@oist.jp}

\author{Francesco Viola}
\affiliation{
 Gran Sasso Science Institute, Viale F. Crispi 7, 67100, L’Aquila, Italy
}

\author{Andrea Mazzino}
\affiliation{
 Department of Civil, Chemical and Environmental Engineering (DICCA), University of Genova, Via Montallegro 1, 16145 Genova, Italy
}
\affiliation{
 INFN, Genova Section, Via Montallegro 1, 16145 Genova, Italy
}

\author{Marco E. Rosti}
\affiliation{
 Complex Fluids and Flows Unit, Okinawa Institute of Science and Technology Graduate University, 1919-1 Tancha, Onna-son, Okinawa 904-0495, Japan
}


\date{\today}

\begin{abstract}
A fully-resolved direct-numerical-simulation (DNS) approach for investigating flexible bodies \added{forced} by a turbulent incoming flow is designed to study the flapping motion of a flexible flag at moderate Reynolds number. The incoming turbulent flow is generated by placing a passive grid at the inlet of the numerical domain and the turbulence level of the flow \added{impacting} the flag can be controlled by changing its downstream distance from the grid. The computational framework is based on the immersed boundary method for dealing with arbitrary geometries and implemented using a graphics-processing-unit (GPU) accelerated parallelisation to increase the computational efficiency. The grid-induced turbulent flow is first characterised by means of \deleted{the} comparison with well-known results for decaying turbulence and a scale-by-scale analysis. Then, the flag-in-the-wind problem is revisited by exploring the effect of the turbulence intensity on self-sustained flapping. Whilst the latter is still manifesting under strong fluctuations, the main features of the oscillation (including its amplitude and frequency) are altered by turbulence, whose fingerprint can also be qualitatively detected by spectral analysis. Besides their relevance for advancing the fundamental understanding of fluid-structure interaction in turbulence, these findings have potential impact for related applications, e.g., aeroelastic energy harvesting or flow control techniques.
\end{abstract}


\maketitle 

\section{Introduction}

\begin{figure*}
\centering
\includegraphics[width=.75\textwidth]{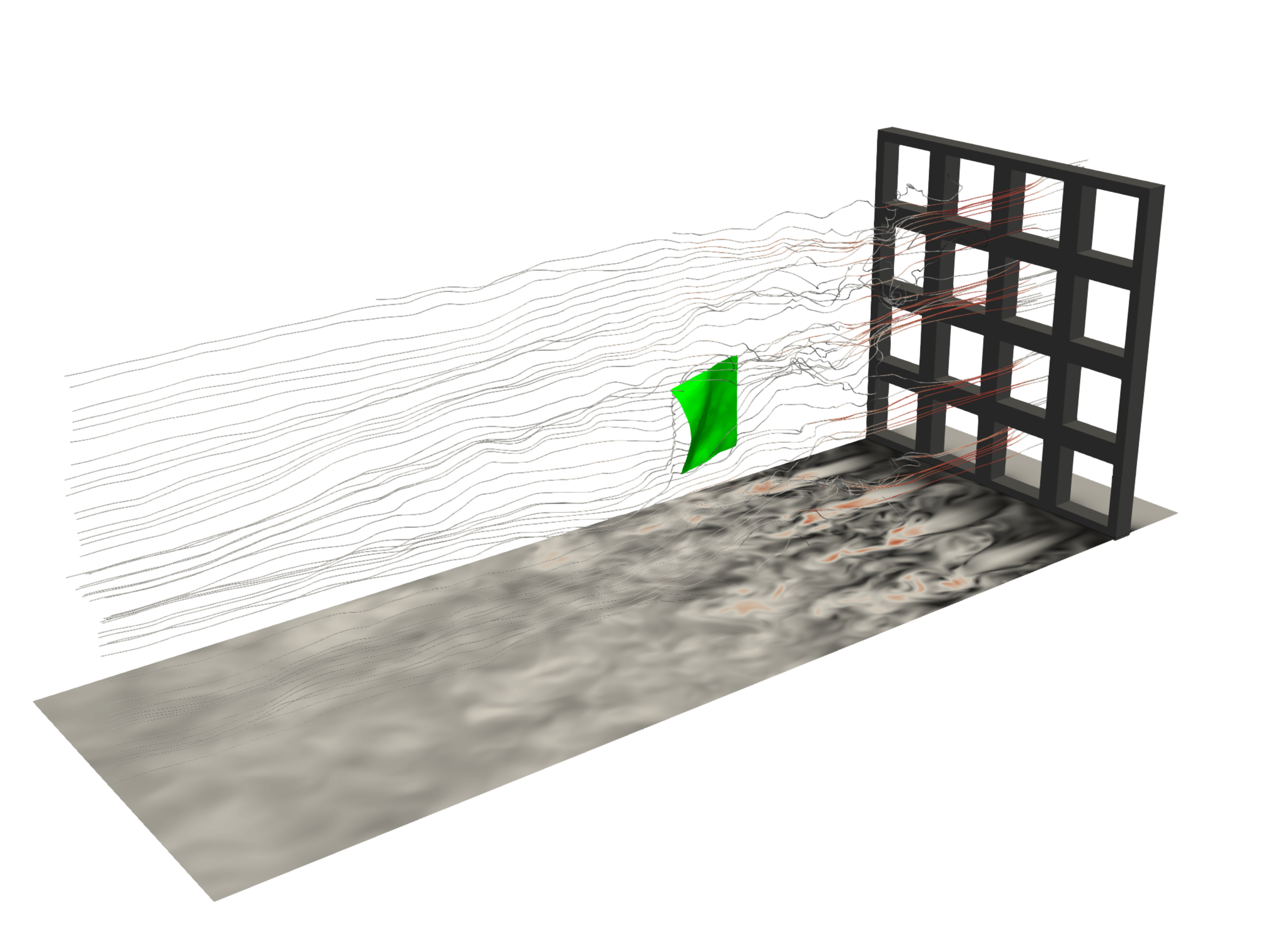}
\caption{
Fluid-structure interaction between a flexible flag (green) and the turbulent flow generated using a passive grid (gray). Snapshot from DNS where the instantaneous fluid flow is shown by means of streamlines in the domain and contours onto a longitudinal plane (normal to the transverse direction), both coloured (from black to red) by the fluid velocity magnitude. The flag is hinged at its leading edge while the other edges are free to move.
}
\label{fig:snapshotFlag1}
\end{figure*}

Fluid-structure interaction (FSI) is ubiquitous and governs many physical problems with relevance in natural processes~\cite{leclercq2018reconfiguration,loiseau2020active,tschisgale2021large} and engineering applications\added{~\cite{hobeck2012artificial,orrego2017harvesting,gallegos2017flags,boccalero2017power,rosti2016direct,rosti2018passive,olivieri2019constructive}}. A formidable and \added{recently increased} effort has been devoted to understand these problems in more detail and, in particular, characterise the aeroelastic instabilities leading to self-sustained, flow-induced oscillations\added{~\cite{mccarthy2016fluttering,olivieri2017fluttering,yu2019review,jin2019flow}}. Notwithstanding, the consolidated knowledge is almost exclusively limited to the case where the incoming flow is laminar. Indeed, only a few studies have considered the role of free-stream turbulence in altering the resulting dynamics~\cite{so2008free,zhu2009turbulence,jin2016spectral}. To fill the gap in the current knowledge, the analysis has therefore to be enriched by considering the typical, yet scarcely explored, configuration where the fluid flow is intrinsically turbulent. In this work, we revisit a well-known FSI problem, i.e. the flapping instability of a flexible flag, in the presence of a turbulent flow generated by using a passive grid (Fig.~\ref{fig:snapshotFlag1}). 

Grid-induced turbulence is one of the most typical configurations adopted for experimental fundamental research on turbulent flows~\citep{george1992decay,krogstad2010grid,davidson2011minimum} and many related problems, e.g. mixing processes~\citep{warhaft1978experimental,sreenivasan1980temperature,zhou2000transport} or aerodynamic testing~\citep{tamura1999effect,vita2018generating}. \added{The} earliest experimental investigations using such configuration\added{s} date back to Batchelor and Townsend~\cite{batchelor1947kolmogoroff,batchelor1947decay}. Since then, numerous studies of both experimental and theoretical nature have focused on various aspects of decaying turbulence, including the existence of self-preserving solutions~\citep{batchelor1948energy,george1992decay} as well as the influence of the so-called initial (i.e., upstream) conditions on the properties of the generated turbulence~\citep{mohamed1990decay,lavoie2005effect,lavoie2007effects,kurian2009grid}. Although the full validation of a comprehensive theory for decaying turbulence is still a debated topic~\citep{djenidi2015power}, it is widely accepted that the turbulent fluctuations decay along the mean flow direction according to a power law that can be expressed (in the laboratory frame of reference) as
\begin{equation}
  \langle u_i'^2 \rangle \sim (x - x_0)^{-n},
  \label{eq:decaySc}
\end{equation}
where $u_i'$ is the fluctuating part of a generic velocity component, $(x - x_0)$ is a properly adjusted distance from the grid and $n$ is the power-law exponent.  

While grid-induced turbulence is a standard configuration in the experimental practice, the situation is remarkably different in the computational framework and, in particular, for what it \added{is considered} the state of the art of \textit{fully-resolved} (i.e., explicitly modeling the presence of the turbulence-generating grid and its geometrical details), direct numerical simulations (DNSs). Indeed, only a few seminal papers addressed the problem \added{from} this perspective~\citep{djenidi2006lattice,nagata2008direct,ertuncc2010homogeneity,laizet2011dns,suzuki2013direct,laizet2015influence}, while the vast majority of previous studies focused on the simulation of freely-decaying homogeneous turbulence in a triperiodic domain imposing some representative initial fluid flow conditions~\cite{antonia2004similarity}. The simulation of grid turbulence in such a fully-resolved approach requires a particular effort on the computational side because of (at least) two critical requirements: \textit{(i)} to appropriately model the initial or upstream conditions (essentially related to the grid geometry), which are known to influence the properties of the resulting turbulent flow~\cite{lavoie2007effects}; \textit{(ii)} to consider a very large extension of the computational domain (in particular along the streamwise direction), in order to describe the whole fully-developed range of turbulence decay. While the first issue can be solved using suitable numerical methods, the second represents an unavoidable challenge that has to be faced by means of an efficient high-performance-computing (HPC) strategy.

Among the first investigations of grid-induced turbulence by fully-resolved numerical simulations, \citet{djenidi2006lattice} originally tackled the problem (using a Lattice-Boltzmann method complemented by a LES correction scheme) modeling the grid as an array of floating squares normal to the incoming flow. Along with highlighting the most delicate aspects in carrying out this kind of computationally-challenging simulation, this study provided a first set of numerical results which are compared with similar experimental measurements. \citet{nagata2008direct} later investigated the turbulent heat transfer considering a thermal mixing layer and investigating several types of grid geometries by means of an immersed boundary method to model the no-slip condition on the grid elements.  Using a similar approach,~\citet{laizet2011dns} and~\citet{suzuki2013direct} performed DNS studies on so-called fractal grids, with the goal of characterising the flow obtained using this specific type of geometry and mostly focusing on the production region relatively close to the grid. 
Overall, these contributions provided valuable knowledge and demonstrated the feasibility of DNS using a fully-resolved approach. Nevertheless, the development and employment of this kind of computational tool is still in its early stage, both for fundamental research and related applications, e.g., in aerodynamics or fluid-structure interaction problems. In particular, the critical aspects limiting the current state of the art are typically represented by the highest Reynolds number and maximum streamwise extension that can be achieved.
 
In  FSI, the effect of turbulence can be crucial on triggering (or suppressing) the onset of aeroelastic instabilities and more generally altering the characteristics of the resulting flow-induced oscillations, thus affecting, e.g., the potential for energy harvesting~\cite{mccarthy2016fluttering,olivieri2019constructive,olivieri2020phd}.
In particular, here we focus on the three-dimensional flapping dynamics of a flexible flag, a problem attracting significant attention both from the fundamental and applied perspective\added{~\cite{huang2010three,xia2015fluid,gallegos2017flags,orrego2017harvesting,yu2019review,rips2020heat,deng2021fluid,kumar2021dynamics}}, however so far substantially limited to the situation where the incoming flow is laminar.
To tackle the problem, we first present an efficient and accurate computational framework for performing fully-resolved DNS of grid turbulence at moderate Reynolds numbers. Then, we revisit the classical flag-in-the-wind problem focusing on the influence of the incoming turbulence, unraveling the main similarities and differences with respect to the laminar case.

The rest of the paper is organised as follows. 
In Sec.~\ref{sec:method}, we first introduce the governing equations and relevant physical parameters (Sec.~\ref{sec:pb}) and then describe the computational methodology and discuss some technical features (Sec.~\ref{sec:num}). Sec.~\ref{sec:flow} concerns the characterisation of the generated turbulent flow and Sec.~\ref{sec:flag} reports our findings on the flag-in-the-wind problem in the case of incoming turbulent flow. Finally, Sec.~\ref{sec:conclusions} draws some conclusions.

\section{Methodology}
\label{sec:method}

\subsection{Problem definition}
\label{sec:pb}

\begin{figure*}
\centering
\includegraphics[width=0.95\textwidth]{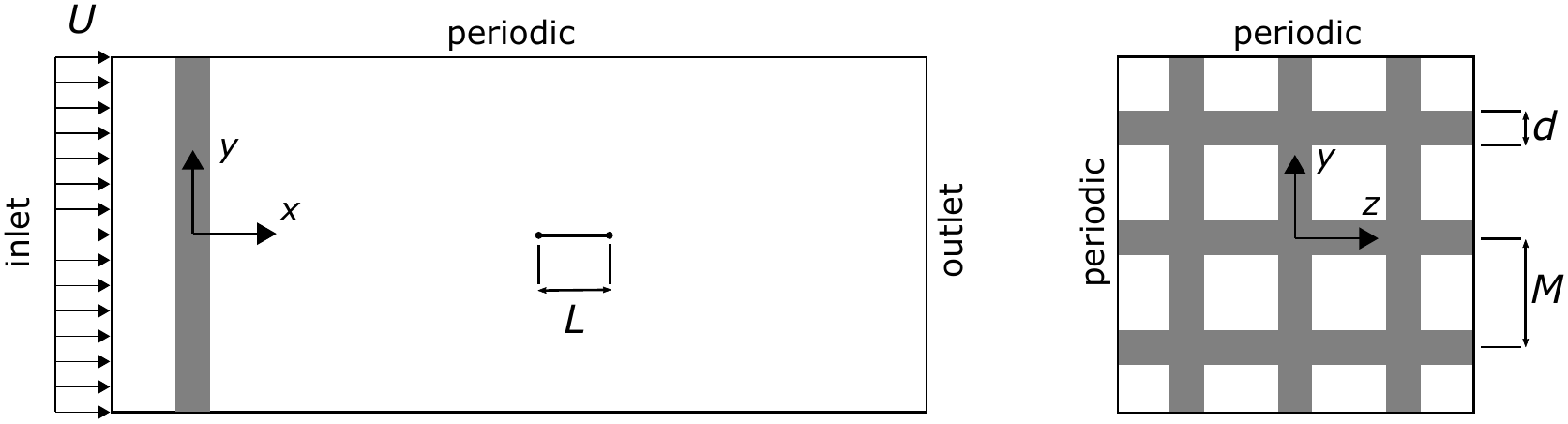}
\caption{
Sketch of the problem (left: sideview, right: front view). A rectangular fluid domain is considered with inlet/outlet conditions in the streamwise direction $x$ and periodic boundary conditions along the transverse directions $y$ and $z$. \added{A regular passive grid used to generate the turbulent flow is placed at a distance of $M/2$ from the inlet plane and characterised by square rods of side $d$ and spacing $M$.} A flexible flag of side $L$ (both in the streamwise and spanwise extension) is placed downstream of the grid.
}
\label{fig:pb-sketch}
\end{figure*}

We consider a Newtonian fluid flow governed by the incompressible Navier-Stokes equations: 
\begin{align}
  \frac{\partial \mathbf{u}}{\partial t} + \mathbf{u} \cdot \nabla \mathbf{u} &= -  \nabla p / \rho + \nu \nabla^2 \mathbf{u} + \mathbf{f}
   \label{eq:NS1}
,\\
  \nabla \cdot \mathbf{u} &= 0,
 \label{eq:NS2}
\end{align}
where $\mathbf{u}=\mathbf{u}(\mathbf{x},t)$ is the velocity and  $p=p(\mathbf{x},t)$ the pressure fields, $\rho$ and $\nu$ are the fluid density and kinematic viscosity, respectively, and $\mathbf{f}=\mathbf{f}(\mathbf{x},t)$ is a body force mimicking the presence of solid bodies (as later discussed in Sec.~\ref{sec:num}). The fluid domain is a box of size $L_x \times L_y \times L_z$ along the streamwise, vertical and spanwise directions, with the origin of the reference frame coinciding with the geometrical centre of the grid used to generate turbulence.  A schematic description of the geometrical setup is shown in Fig.~\ref{fig:pb-sketch}. A uniform fluid velocity $\mathbf{u} = (U,0,0)$ is imposed at the inlet, whereas a non-reflecting convective boundary condition is used at the outlet. Periodic boundary conditions apply along the vertical and spanwise directions, this setup being representative of the central region of a wind-tunnel sufficiently far from the surrounding walls.

\added{In order to induce a well characterized turbulent flow, a monoplane, passive grid is placed at a streamwise distance of $M/2$ from the inlet plane, as sketched in Fig.~\ref{fig:pb-sketch}. The no-slip boundary condition applies at the solid surface of the grid. The grid geometry is characterised by the spacing $M$ between the elements and by their cross-sectional shape and size. Here, we consider a square cross section with rod side $d$.
Note that we choose a regular, monoplane grid (i.e., the vertical and horizontal rods lie in the same plane without any offset along the streamwise direction),} but other configurations could be considered as well, including active grids using appropriate boundary conditions and geometrical elements. The problem is made nondimensional using the grid spacing $M$ and the unperturbed flow velocity $U$, so that we define the nominal Reynolds number $\mathrm{Re}_M = {U M}/{\nu}$ and the geometrical ratio $M/d$ or, equivalently, the grid porosity $\alpha = (1-d/M)^2$ and solidity $\sigma = 1-\alpha$, as the main nondimensional governing parameters. Here, we fix $M/d=4$, or equivalently $\sigma=44\%$, in order to achieve a rather strong turbulence intensity while ensuring a good isotropy~\cite{kurian2009grid}. Moreover, such configuration will allow \added{comparison of} our results with available experimental data from~\citet{djenidi2015power}.

As a representative FSI, we consider the well-known (three-dimensional) problem of a flag flapping in the wind~\cite{huang2010three}. A square flag of side $L$ is placed at a given streamwise distance downstream of the grid (and centered in the middle of the transverse plane). We assume $L=M$ so that the integral lengthscale of the incoming turbulent flow is strictly comparable with the size of the elastic objects. From this choice it also follows that the flag and grid nominal Reynolds numbers match, i.e. $\mathrm{Re}_L = U L / \nu = \mathrm{Re}_M = \mathrm{Re}$. Besides this parameter, in the case of a uniform incoming flow, the flapping dynamics is controlled by the mass ratio $\rho = (\rho_\mathrm{s}/\rho_\mathrm{f}) \, (h/L)$ (where $\rho_\mathrm{s}$ is the solid density and $h$ is the flag thickness) and the bending coefficient $\beta = B / (\rho_\mathrm{f} U^2 L^3)$, where $B$ is the dimensional bending stiffness. In order to compare with available results in the laminar case~\cite{huang2010three,tian2014fluid,lee2015discrete,de2016moving}, we neglect the presence of gravity and set $\rho =1$ and $\beta = 10^{-4}$; the flag is always initialised in the straight configuration with a pitching angle $\theta = 0.1 \pi$ to trigger the flapping instability.

\subsection{Numerical technique}
\label{sec:num}

A critical aspect in the numerical simulation of grid-induced turbulence is represented by the streamwise extension needed to capture the whole dynamical range of interest, i.e., from the production region to that of the final decay~\citep{djenidi2006lattice}. At the same time, the spatial resolution needs to be sufficient to accurately describe the flow features especially in the vicinity of the solid surfaces. The combination of these requirements leads to a dramatic increase in the total number of computational nodes. To tackle such a challenging problem, we employ the recently developed GPU-based parallel version of the AFiD solver~\citep{zhu2018afid}. Hereafter, we briefly discuss its main features, while for additional information the reader is referred to Refs.~\cite{verzicco1996finite,zhu2018afid,viola2020fluid}.

Eqs.~\eqref{eq:NS1} and~\eqref{eq:NS2} are solved numerically using the fractional-step method on a regular staggered Eulerian mesh. The spatial discretisation is performed using the second-order centered finite-difference method. The nonlinear terms are discretised in time using the second-order Adams-Bashfort\added{h} scheme whereas the diffusive terms are computed with the second-order Crank-Nicolson scheme. The no-slip condition on the wet surfaces of the immersed solid bodies is imposed using the immersed boundary (IB) method~\citep{fadlun2000combined,de2016moving}, with the solid bodies discretised onto a non-conformal triangular mesh. A provisional, non-solenoidal velocity field is first computed using an approximate factorization technique and then updated by applying the IB forcing. Then, the provisional velocity is projected onto a divergence-free space after solving the resulting Poisson equation enforcing mass conservation.

Two different types of direct-forcing IB methods are used to enforce the no-slip boundary condition between the fluid and the solid surfaces: \textit{(i)} for the fixed surface of the grid used to generate turbulence, we employ the Eulerian method proposed in Ref.~\cite{fadlun2000combined}; \textit{(ii)} for the moving surface of the flapping flag, we employ a Lagrangian approach based on a moving-least-squares method~\cite{vanella2009moving,de2016moving}. Indeed, the first method is computationally less demanding and ensures good accuracy for fixed bodies (such as the passive grids here considered) as well as for bodies with prescribed motion (thus being potentially used for the simulation of active grids). Instead, when the dynamics is fully governed by the resulting fluid-structure interaction (e.g., the flapping flag), the second method, which is based on a moving-least-squares (MLS) interpolation, is more effective in minimising the spurious numerical oscillations~\cite{vanella2009moving}. Both methods have been extensively validated and used for multiphase and biological FSI problems~\citep{fadlun2000combined,verzicco2000large,de2016moving,spandan2017parallel,olivieri2019constructive,viola2020fluid}.

\begin{figure*}
\centering
\includegraphics[width=0.8\textwidth]{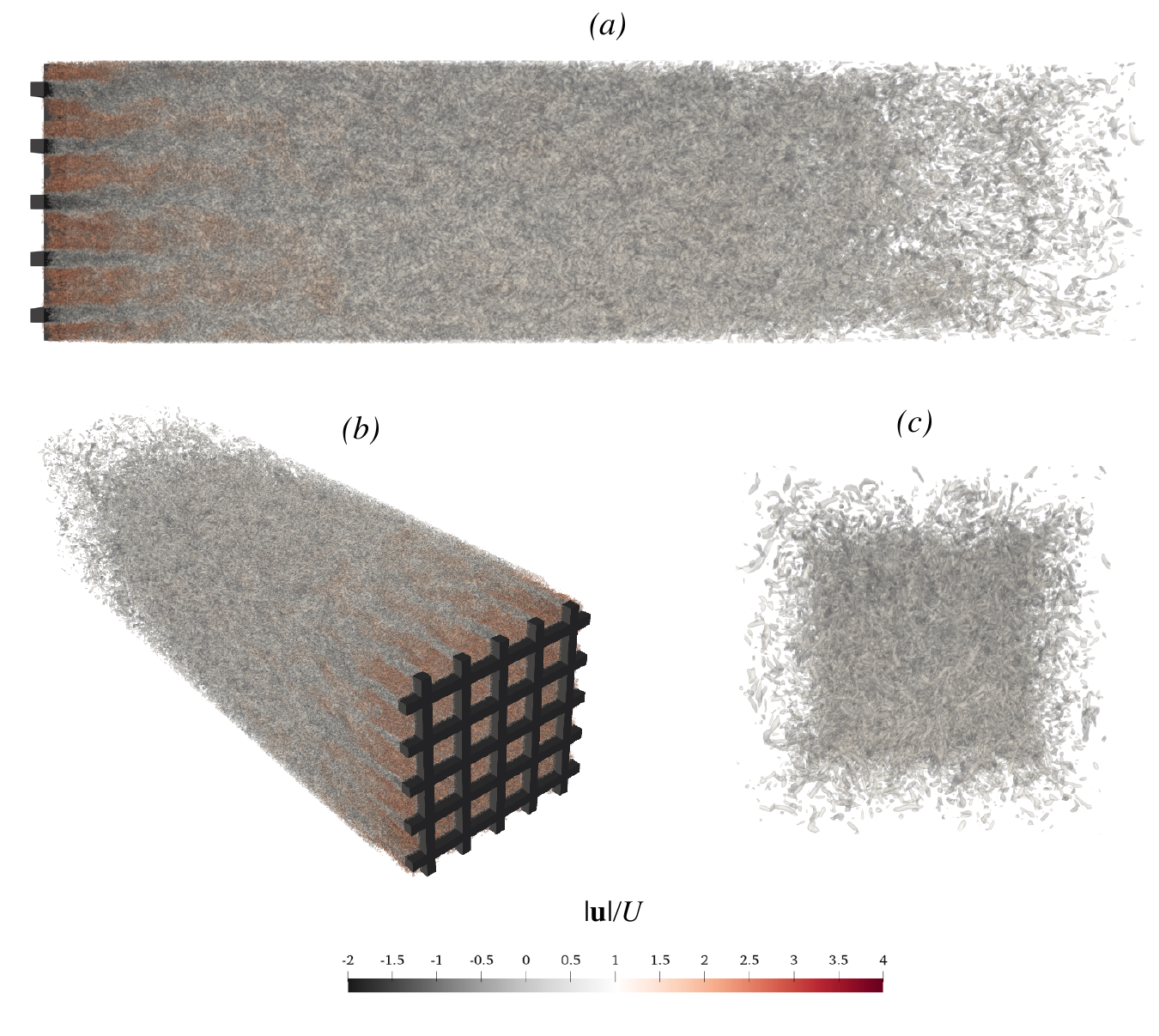}
\caption{
Snapshots from DNS of grid-induced turbulence at $\mathrm{Re}_M=5000$: \textit{(a)} sideview, \textit{(b)} 3-D view and \textit{(c)} backview. The figure shows the instantaneous vortical structures detected using the $Q$-criterion and coloured by the streamwise fluid velocity (from black to red).}
\label{fig:snapshotSide}
\end{figure*}

A two-dimensional spring-network structural model is used for simulating the deformable bodies~\cite{de2016moving}. The flag is modeled as a thin elastic membrane and discretised onto a Lagrangian mesh made of triangular elements. The mass is distributed on the vertices of the triangles which are connected by different types of springs to model both the in-plane and out-plane deformation~\cite{fedosov2010systematic}. In our case, we set the spring coefficients so that the flag is essentially inextensible and the only parameter controlling the elasticity is the bending stiffness $B$. 
Note that, as the Eulerian grid for the fluid flow is refined, the Lagrangian resolution has to be increased as well to ensure the correct enforcement of the no-slip boundary condition. \added{On the other hand, increasing the Lagrangian resolution would also reflect on a smaller simulation timestep required by the structural solver.
To decouple these two numerical constraints (the first arising from the fluid-solid coupling and the second specifically associated with the structural dynamics), we proceed as follows. We set the (base) Lagrangian resolution to satisfy the latter, therefore retaining the same mesh independent of the Eulerian grid resolution and consequently the same required timestep. For dealing with the former, we employ an adaptive Lagrangian tiling procedure~\cite{viola2020fluid} so that each triangular element is successively refined in order to match the local Eulerian resolution. Such procedure can be performed just once at the start of the simulation or dynamically at every timestep and is based on the combination of quadratic and baricentric tiling rules. Overall, the advantage of using the proposed strategy is the decreased number of Lagrangian elements along with a larger simulation timestep.
}

A final note regards the code implementation and parallel computing performance. In this work, we employ the porting of the existing CPU version (using hybrid MPI/OpenMP parallelisation) to GPU-accelerated architectures using the CUDA Fortran library and an extensive use of CUF kernel directives~\cite{ruetsch2013cuda,zhu2018afid}. A (one-dimensional) slab domain decomposition is used to distribute the simulation over four NVIDIA Tesla V100 GPUs using the Saion cluster at OIST. In the baseline configurations explained in the following sections, the computing time per iteration is about $0.075 \, \mathrm{s}$ for the characterisation of grid-induced turbulence (Sec.~\ref{sec:flow}) and $0.13 \, \mathrm{s}$  for the complete simulation of the FSI problem (Sec.~\ref{sec:flag}).

\section{Characterisation of grid-induced turbulence}
\label{sec:flow}

As a first step, we focus on characterising the turbulent flow that is generated by the presence of the fixed grid. To this aim, we maximise the streamwise domain extension to $L_x = 120 M$, whereas for the transverse (vertical and spanwise) directions a much smaller extension of $L_y = L_z = 5M$ is used. The simulations are carried out on a uniform Eulerian mesh made of $3072\times128\times128$ nodes using a fixed timestep $\Delta t U/M = 5 \times 10^{-4}$. Results on the convergence with respect to the mesh resolution and transverse domain extension are reported in Appendix~\ref{sec:convergence}, showing only a minimal influence on the main quantities of interest.

The study is performed considering two different Reynolds numbers: \textit{(i)} $\mathrm{Re}_M=1000$ that will be the configuration used for the complete FSI simulation (see next Section~\ref{sec:flag}) and \textit{(ii)} $\mathrm{Re}_M=5000$ that will be considered here for a more detailed analysis and to perform a comparison with the experimental data of Ref.~\cite{djenidi2015power}.
\added{We disregard the initial transient and start accumulating the statistics after $50 \, M/U$ (nondimensional) time units, having verified (both by probing and visualising the flow) that the turbulence is fully developed throughout the domain. The simulations are advanced in time up to $t\,U/M = 2000$ in order to achieve statistical convergence (with the maximum statistical error of about 2\% when halving the sampling interval, measured for the quantities with slower convergence such as the large-scale anisotropy ratio).} The quantities of interest are computed by sampling the flow time history with a stencil of $2151\times3\times3$ virtual probes arranged around the centerline (with a uniform spacing of $M/2$ in the streamwise direction and $2M/3$ in the transverse directions) and making use of Taylor's frozen turbulence hypothesis. Additionally,  at $\mathrm{Re}_M=5000$ we extract a number of slices in the transverse homogeneous plane (which are evenly-spaced in the streamwise direction by $5M$) for computing the energy spectra along with the same aforementioned quantities.

\begin{figure}
\centering
\includegraphics{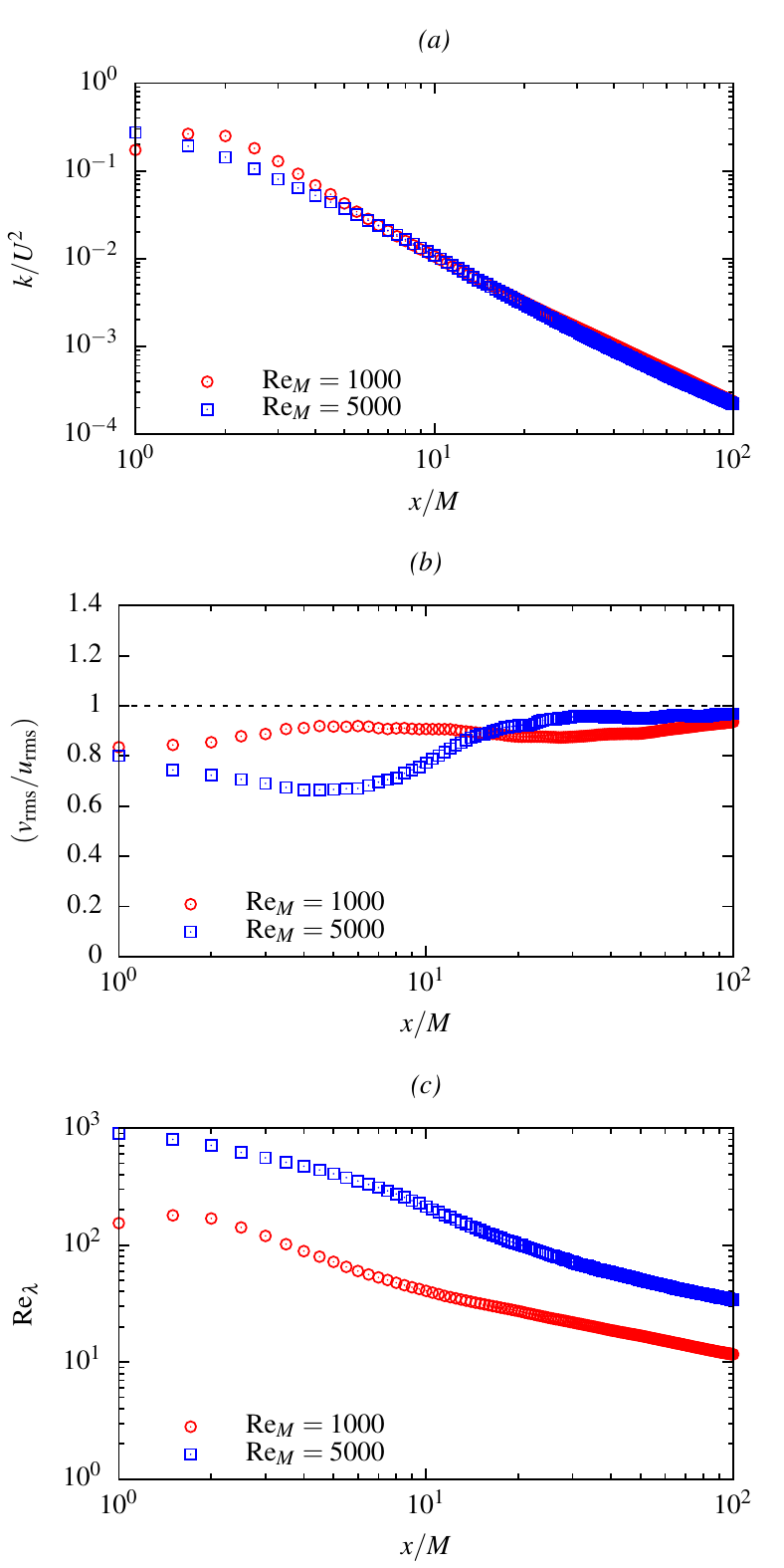}
\caption{
Grid-induced turbulence main quantities as a function of the normalized streamwise coordinate: \textit{(a)} turbulent kinetic energy; \textit{(b)}  large-scale anisotropy; \textit{(c)} Reynolds number based on the Taylor lengthscale. Two different Reynolds numbers are tested: $\mathrm{Re} = 1000$ (red circles) and $5000$ (blue squares).
}
\label{fig:decayGeoRe}
\end{figure}

Fig.~\ref{fig:snapshotSide} provides a snapshot from the simulation at $\mathrm{Re}_M=5000$. For a better visualisation close to the grid, only the region up to $x/M\approx25$ in the streamwise extent is shown instead of the full extent of the domain $x/M\approx 120$. From both the side and 3-D views in Fig.~\ref{fig:snapshotSide}\textit{(a,b}), qualitative insight is given on how turbulence is first generated in the vicinity of the grid and then starts to decay as the streamwise distance further increases. Moreover, it can be observed that the characteristic size of the vortical structures, here detected as the isosurfaces of the positive second invariant of the velocity gradient $Q$~\citep{chakraborty2005relationships}, increases with the distance from the grid. On the other hand, the flow appears to be statistically homogeneous along the $y$ and $z$ direction, Fig.~\ref{fig:snapshotSide}\textit{(c}), as expected. Consequently, when referring to the transverse component we will consider the average between these two directions. 

Fig.~\ref{fig:decayGeoRe} shows the main quantities characterising the turbulent flow as a function of the normalized streamwise coordinate $x/M$: \textit{(a)} the (normalized) turbulent kinetic energy, \textit{(b)} the large-scale flow anisotropy and \textit{(c)} the micro-scale (or turbulent) Reynolds number. The turbulent kinetic energy $k/U^2$ (where $k = \langle u^2 +v^2 +w^2 \rangle / 2$) is shown in Fig.~\ref{fig:decayGeoRe}\textit{(a)}. The characteristic scenario of decaying turbulence can be noticed, so that the range of observation can be roughly divided in two regions: a \textit{production} region (for $x/M \lesssim 10$) where turbulence is developed and a \textit{decay} region (for $x/M \gtrsim 10$) where the turbulent kinetic energy approximately follows the well known power-law decay~\eqref{eq:decaySc}. 
\added{In the latter, the differences observed when varying $\mathrm{Re}_M$ appear rather minimal (although  they may reflect in a subtle variation of the decay exponent $n$~\citep{djenidi2015power}).}
Next, we look at the (large-scale) anisotropy of the flow by considering the ratio between the streamwise and transverse velocity fluctuations $v_\mathrm{rms}/u_\mathrm{rms}$, reported in Fig.~\ref{fig:decayGeoRe}\textit{(b)}. Approaching the decay region (i.e., for $x/M \gtrsim 10$), this quantity appears to be approximately constant in a large portion of the observed range, with $v_\mathrm{rms}/u_\mathrm{rms}$ varying between about 0.8 and 1. A more evident influence of the Reynolds number can be noticed with the flow becoming more isotropic when increasing $\mathrm{Re}_M$, as expected. At $\mathrm{Re}_M=5000$, in particular, an essentially isotropic region is observed for $x/M \gtrsim 30$. Finally, the micro-scale Reynolds number $\mathrm{Re}_\lambda = \lambda_f u_\mathrm{rms} / \nu$, based on the (longitudinal) Taylor lengthscale $\lambda_f$, is shown in Fig.~\ref{fig:decayGeoRe}\textit{(c)}. 
Overall, $\mathrm{Re}_\lambda$ is decreasing with the streamwise coordinate although with a relatively slow decay rate. As expected, a remarkable variation of this quantity is observed when varying $\mathrm{Re}_M$. On the other hand, the difficulty in achieving a large micro-scale Reynolds number, i.e. $\mathrm{Re}_\lambda \gtrsim 10^{2}$, using a classical passive grid is well known~\cite{thormann2014decay}.

\begin{figure*}
\centering
\includegraphics{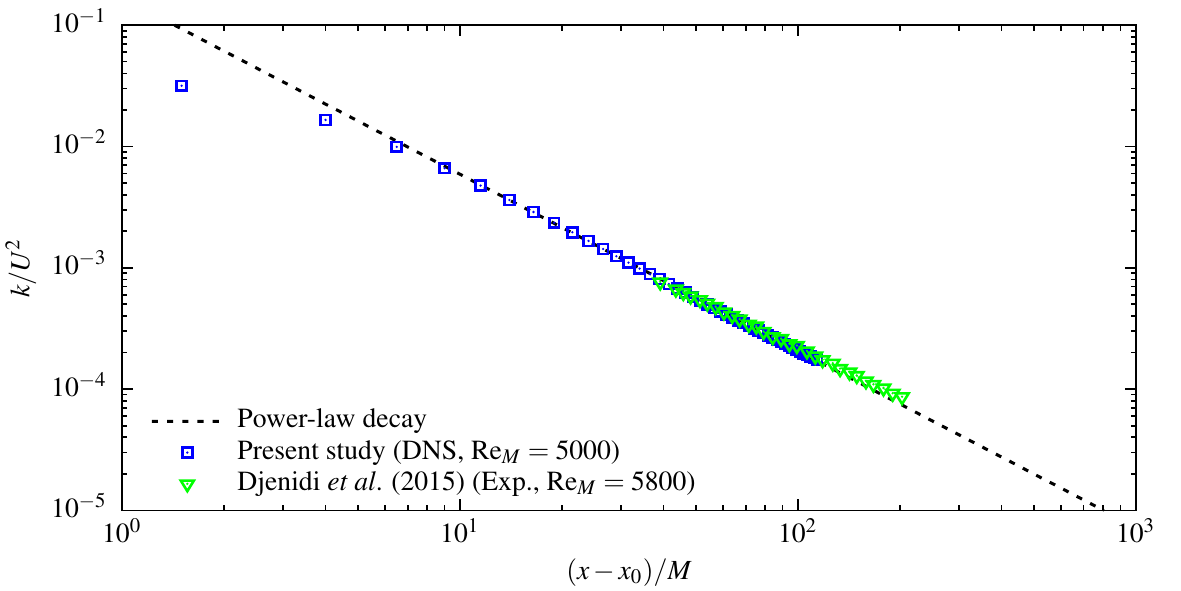}  
\caption{
Comparison of DNS results of turbulent kinetic energy at $\mathrm{Re}_M=5000$ with the experimental measurements of~\citet{djenidi2015power} and power-law fitting, Eq.~\eqref{eq:decay}. Note that the quantities are plotted as a function of the adjusted streamwise distance using the virtual origin $x_0$.
}
\label{fig:decay2}
\end{figure*}

\begin{figure*}[t]
\centering
\includegraphics{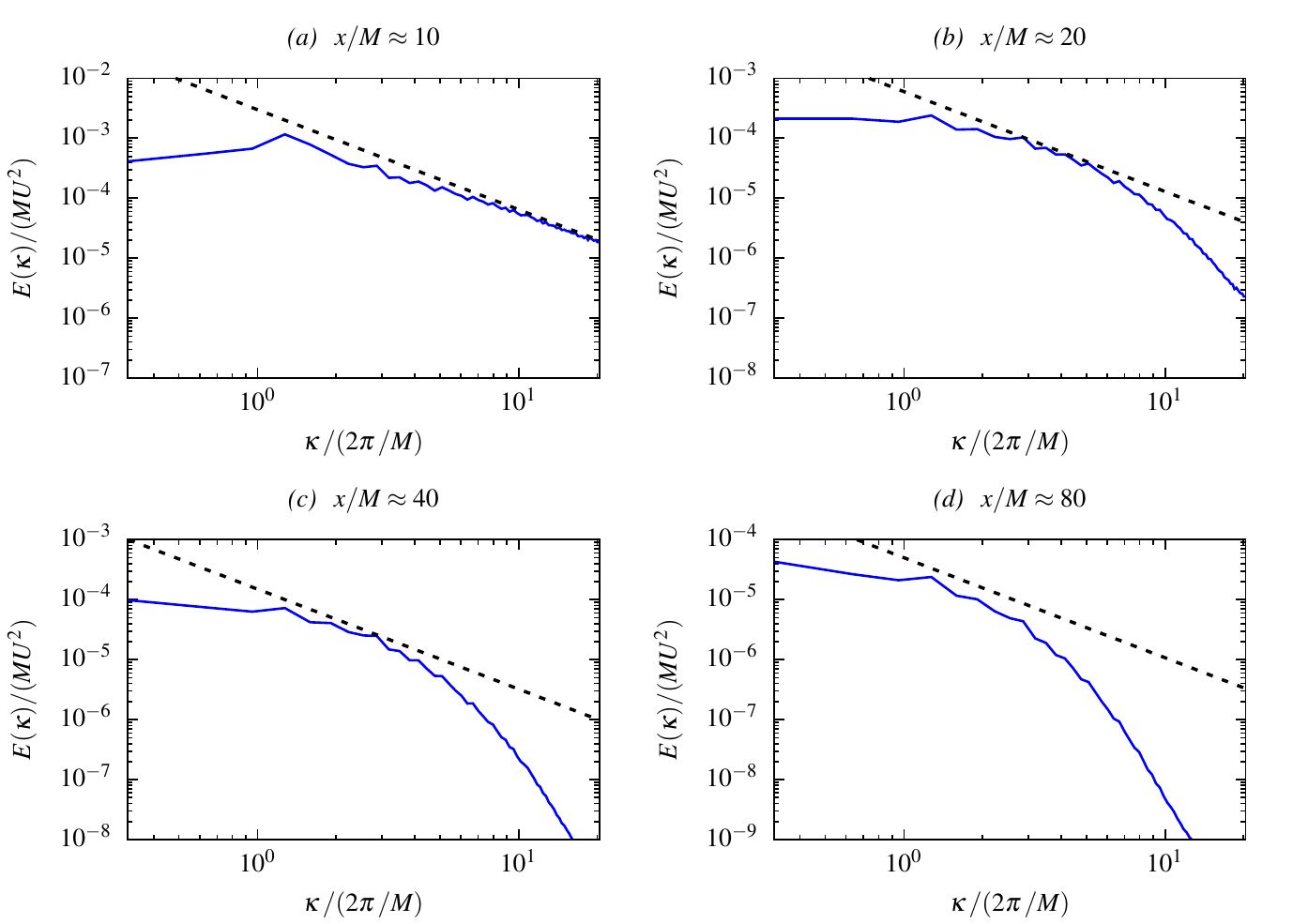} 
\caption{
Energy spectra of turbulent flow at $\mathrm{Re}_M=5000$ at four different streamwise locations from the grid: \textit{(a)}~$x/M\approx10$, \textit{(b)}~$x/M\approx20$, \textit{(c)}~$x/M\approx40$, \textit{(d)}~$x/M\approx80$. The dashed line indicates the characteristic Kolmogorov scaling $\kappa^{-5/3}$ in the inertial subrange.
}
\label{fig:spectra}
\end{figure*}

Our results obtained at $\mathrm{Re}_M=5000$ can be representative of typical grid-induced turbulence conditions and are therefore compared in Fig.~\ref{fig:decay2} with the experimental measurements reported by~\citet{djenidi2015power}. Specifically, we compare with their findings at $\mathrm{Re}_M=5800$ using a perforated grid with square holes having solidity $\sigma \approx 43\%$ (indicated as SSQ43 in Ref.~\cite{djenidi2015power}). Despite the minor discrepancies due to non-negligible difference in the nominal parameters, the numerical data appear to be fully consistent with similar experimental evidence. Furthermore, as shown in Fig.~\ref{fig:decay2}, we quantitatively assess the decay of the turbulent kinetic energy by fitting our data to the well-known power-law expression~\citep{mohamed1990decay}
\begin{equation}
\frac{k}{U^2} = A \,\frac{(x - x_0)}{M}^{-n},
  \label{eq:decay}
\end{equation}
where three free parameters appear: a multiplicative factor $A$, the virtual origin $x_0$ and the decay exponent $n > 0$. The delicate issue of how to perform accurately the fitting procedure has been discussed by several authors, see e.g. Refs.~\cite{mohamed1990decay,lavoie2007effects,kurian2009grid}. Moreover, Ref.~\cite{djenidi2015power} recently revised the applicability of Eq.~\eqref{eq:decay}, suggesting the use of a modified form where the decay exponent is not constant.  Here, we adopt the classical form and perform the fitting procedure as follows: firstly, we deduce the most appropriate value for the virtual origin $x_0$ on a trial-and-error basis by varying this parameter in order to observe the most extended range of power-law decay when plotting $k$, or equivalently the most horizontal plateau when plotting $\lambda^2 / [M (x-x_0)]$, as a function of $(x-x_0)$~\citep{lavoie2007effects}. Then, while keeping fixed the virtual origin, we perform a least-squares fit to obtain the other two parameters, $A$ and $n$. The following values are obtained and used for plotting the power-law in Fig.~\ref{fig:decay2}\textit{(a)}: $A = 0.17$, $x_0/M = 4$ and $n = 1.45$.
Despite the well-known sensitivity of $n$ with respect to the flow properties as well as the details of the measurement method, our data are in good agreement with DNS results reported in the literature~\cite{djenidi2015power}. Lastly, we point out that the slight deviation of the experimental measures shown in Fig.~\ref{fig:decay2} is consistent with the different nominal Reynolds number. 

\begin{figure*}
\centering
\includegraphics{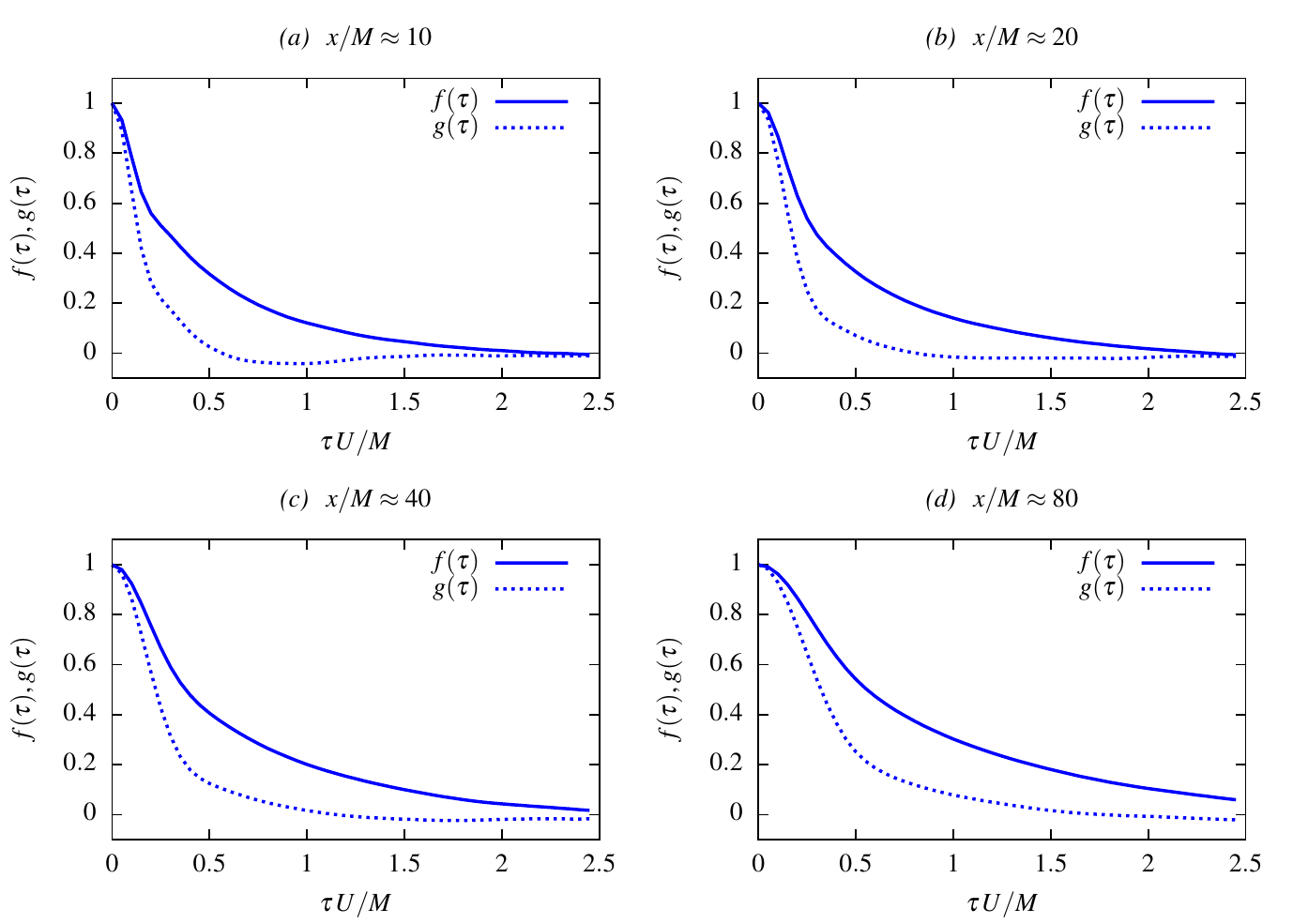}
\caption{
Velocity autocorrelation coefficients of turbulent flow at $\mathrm{Re}_M=5000$ at four different streamwise locations from the grid: \textit{(a)}~$x/M\approx10$, \textit{(b)}~$x/M\approx20$, \textit{(c)}~$x/M\approx40$, \textit{(d)}~$x/M\approx80$. 
The longitudinal and transverse autocorrelation coefficients, defined by  Eqs.~\eqref{eq:fuTau} and~\eqref{eq:guTau}, are denoted by $f(\tau)$ and $g(\tau)$, respectively.
}
\label{fig:autocorr}
\end{figure*}

Fig.~\ref{fig:spectra} shows the turbulent energy spectra $E(\kappa)$ computed at four streamwise locations $x/M \approx \{10,20,40,80\}$. It can be observed that $x/M\approx10$ corresponds to the very early onset of the decay region. The numerical results are compared with the classical Kolmogorov scaling $k^{-5/3}$ that is expected for the inertial subrange. Taking into account the moderate Reynolds number achieved in our configuration, it can be noticed a reasonable agreement with such prediction. 
Comparing the different panels, it can be appreciated how the turbulent kinetic energy (i.e., the integral of the spectrum) progressively decays for increasing $x/M$. In fact, using the corresponding values of the integral lengthscale and turbulent kinetic energy, it is possible to normalise these plots with all curves collapsing on each other within the inertial subrange (not shown).

Fig.~\ref{fig:autocorr} collects the plots of the velocity autocorrelation coefficients at the same four streamwise locations $x/M \approx \{10,20,40,80\}$. We define the longitudinal and transverse autocorrelation coefficients, $f(\tau)$ and $g(\tau)$, respectively as:
\begin{equation}
f (\tau) = \frac { \langle u(\mathbf{x}+\tau U \mathbf{e_x}) \, u(\mathbf{x}) \rangle } { \langle u^2(\mathbf{x}) \rangle },
\label{eq:fuTau}
\end{equation}
\begin{equation}
g (\tau) = \frac { \langle v(\mathbf{x}+\tau U \mathbf{e_x}) \, v(\mathbf{x}) \rangle } { \langle v^2(\mathbf{x}) \rangle } ,
\label{eq:guTau}
\end{equation}
where $\tau$ is the temporal separation. Taylor's frozen turbulence hypothesis is used to relate the temporal and spatial separation, i.e. we assume $r_x \approx \tau U$. The validity of such assumption has been directly verified by evaluating the equality $\partial_t u + U \partial_x u\approx 0$ throughout the fluid domain. The autocorrelation coefficients are used to compute the characteristic lengthscales of the turbulent flow. Looking at Fig.~\ref{fig:autocorr}, the autocorrelation is essentially going to zero for $\tau U / M \approx 2.5$, thus confirming that the transverse domain extension is sufficient to describe the whole range of scales of fluid motion.

\begin{figure*}
\centering
\includegraphics{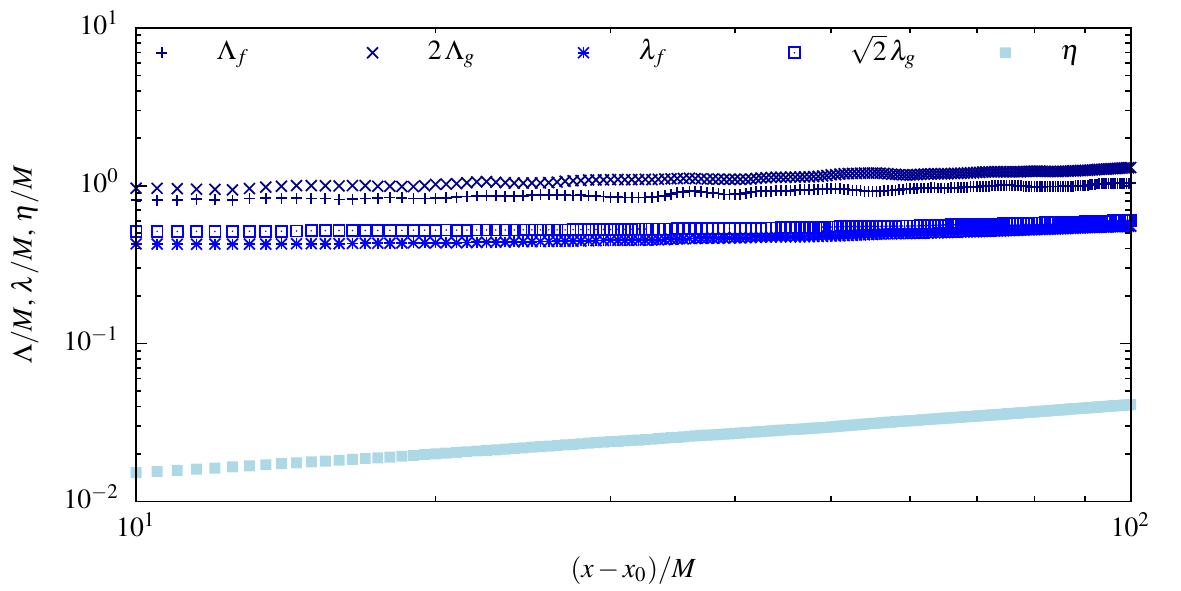}
\caption{
Overview of characteristic, i.e., integral (dark blue), Taylor (blue) and Kolmogorov (light blue), lengthscales at $\mathrm{Re}_M=5000$. The transverse (integral and Taylor) lengthscales are multiplied to match the longitudinal ones by the corresponding factors predicted in the case of isotropic turbulence.
}
\label{fig:charLen}
\end{figure*}

An overview of how the characteristic lengthscales of the turbulent flow  (i.e. integral, Taylor and Kolmogorov scales) evolve approximately within the decay region is given in Fig.~\ref{fig:charLen}. Such lengthscales are obtained from the autocorrelation coefficients and properly defined in the following.
Firstly, the integral lengthscales $\Lambda$ are representative of the largest turbulent eddies and are computed by integrating the corresponding autocorrelation coefficient over the range of separation.
In Fig.~\ref{fig:charLen}, the transverse scale is multiplied by 2 in order to graphically corroborate the prediction $\Lambda_f = 2 \Lambda_g$ obtained for isotropic turbulence~\citep{pope2000turbulent} (where $\Lambda_f$ and $\Lambda_g$ are the longitudinal and transverse integral lengthscales, respectively). It is also evident that $\Lambda_f \approx 2 \Lambda_g \approx M$, i.e. the integral lengthscale is substantially comparable with the grid spacing, although it progressively increases with $x/M$. Nevertheless, this further confirms that the chosen domain is adequate to correctly capture the largest existing flow structures. Next, the Taylor lengthscales $\lambda$ representative of intermediate-size eddies are obtained after a Taylor series expansion of the corresponding autocorrelation coefficient near the origin~\citep{pope2000turbulent}. The theoretical background for isotropic turbulence yields $\lambda_f / \lambda_g = \sqrt{2}$. In Fig.~\ref{fig:charLen} we account for this factor to relate the longitudinal and transverse scales, showing very good agreement especially when approaching the region where the flow becomes more isotropic (see Fig.~\ref{fig:decayGeoRe}\textit{(b)}). Lastly, we focus on the smallest flow scales dominated by viscous dissipation. We compute the energy dissipation rate as $\epsilon =  30 \nu \, { u^2_\mathrm{rms} } /{ \lambda_f^2 } = 15 \nu \, { u^2_\mathrm{rms} } /{ \lambda_g^2 }$ and obtain the Kolmogorov lengthscale $\eta = \left( {\nu^3}/{\epsilon} \right)^{1/4}$~\citep{tropea2007springer}. The evolution of the latter is depicted in~Fig.~\ref{fig:charLen}, showing a tendency to grow with the streamwise coordinate as the turbulent eddies decay and diffuse similar to the other characteristic scales (although according to different scaling laws~\cite{kurian2009grid}).

\section{Flapping flag in grid turbulence}
\label{sec:flag}

In this Section, the flag-in-the-wind FSI problem is investigated in the case of a turbulent incoming flow. The setup is depicted in Fig.~\ref{fig:snapshotFlag1} where the grid-induced turbulence is \added{impacting} the flag and perturbing the flapping motion that would naturally arise in the laminar case.

To control the turbulent intensity $\mathit{Tu} = \sqrt{\frac{2}{3} k} / U $ in the proximity of the flag, we vary the distance from the grid $X_\mathrm{LE}/ M$ at which the leading edge (LE) is placed. Two distances $X_\mathrm{LE} / M = \{ 4, 8\}$ that correspond to $\mathit{Tu} \approx \{ 40\%, 20\%\}$ are tested. Additionally, the simulation at $\mathit{Tu}=0$ (i.e., $X_\mathrm{LE}/M \to \infty $) is performed by removing the grid. After validating our numerical method (see Appendix~\ref{sec:convergence}), the analysis is performed at $\mathrm{Re}_L = 1000$ to achieve a configuration where the incoming flow shows the essential features of grid turbulence, as shown in Sec.~\ref{sec:flow}. The simulations are carried out on a uniform mesh made of $900\times300\times300$ nodes using a fixed timestep $\Delta t U/M = 5 \times 10^{-4}$, reaching a final time of $t \, U/M = 300$. Specifically, the domain extension is set to $L_x = 12 M$ and  $L_y = L_z  = 4 M$  whereas the grid resolution is improved in order to better describe the flow details close to the flag and consequently the coupled flapping dynamics. Note that our focus is on strong turbulent fluctuations that are achieved by placing the flag at a relatively short distance from the grid, but sufficiently far from the highly inhomogeneous region immediately downstream  of the grid.

\begin{figure*}
\centering
\includegraphics{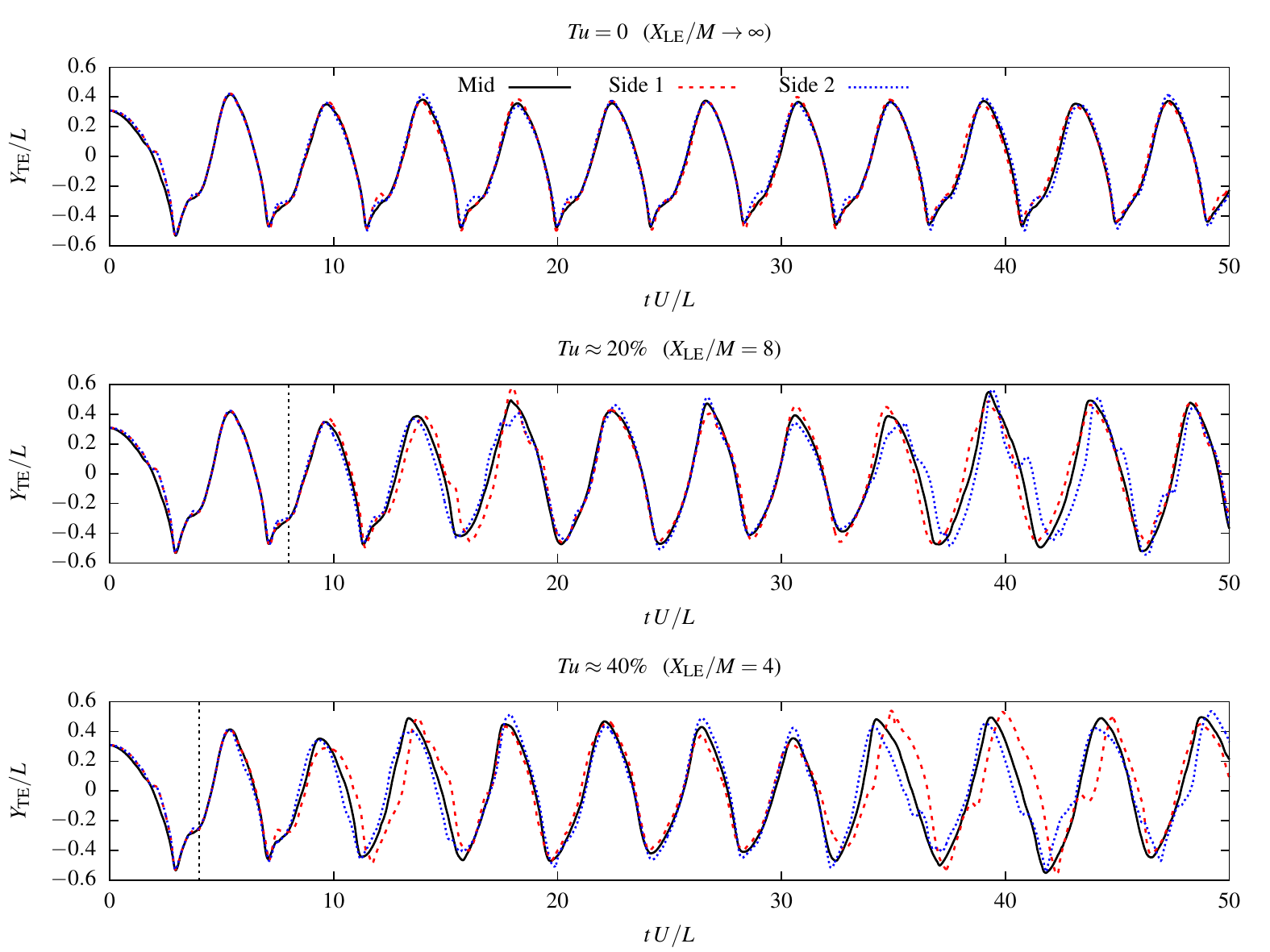}
\caption{
Time history of trailing edge middle (solid black) and side (dashed red and dotted blue) points of flapping flag at different turbulence intensities. The plots show the initial stage of the simulation. For $\mathit{Tu} \approx \{ 20\%, 40\%\}$ he vertical dashed line indicates the time $X_\mathrm{LE}/U$ at which the turbulent front reaches the flag.
}
\label{fig:cfrTurb_TE}
\end{figure*}

Fig.~\ref{fig:cfrTurb_TE} collects the time history of the trailing edge (TE) transverse motion for each of the investigated configurations. It is worth noticing a close resemblance of the results in the early stage of the simulation (i.e., approximately within the first flapping cycle). As graphically shown in the corresponding plots, this is simply explained by the fact that for $\mathit{Tu} \approx \{ 20\%, 40\%\}$ the turbulent front has not reached the flag yet. The flapping motion is therefore always triggered in the same way as the laminar case. However, after few flapping cycles a deviation of the turbulent cases from the latter becomes evident, i.e., when the perturbation exerted by the incoming grid turbulence is effective. As a result, the oscillation becomes less regular with a clear variation of the peak-to-peak amplitude for increasing $\mathit{Tu}$. On the other hand, it has to be pointed out that the aeroelastic instability appears to be robust, thus remaining the main mechanism driving the self-sustained motion of the flag. For a deeper understanding, Fig.~\ref{fig:cfrTurb_TE} shows the time history for both the TE side and mid points. In the laminar case, the differences between these observables are negligible, thus suggesting that the motion is essentially two-dimensional with no significant activation modes in the spanwise direction. Conversely, this is not the case in the presence of incoming turbulence, as indicated by the shift between the side and mid point time traces. Such effect is clearly more pronounced when increasing the turbulence intensity.

\begin{figure*}
\centering
\includegraphics[width=1.0\textwidth]{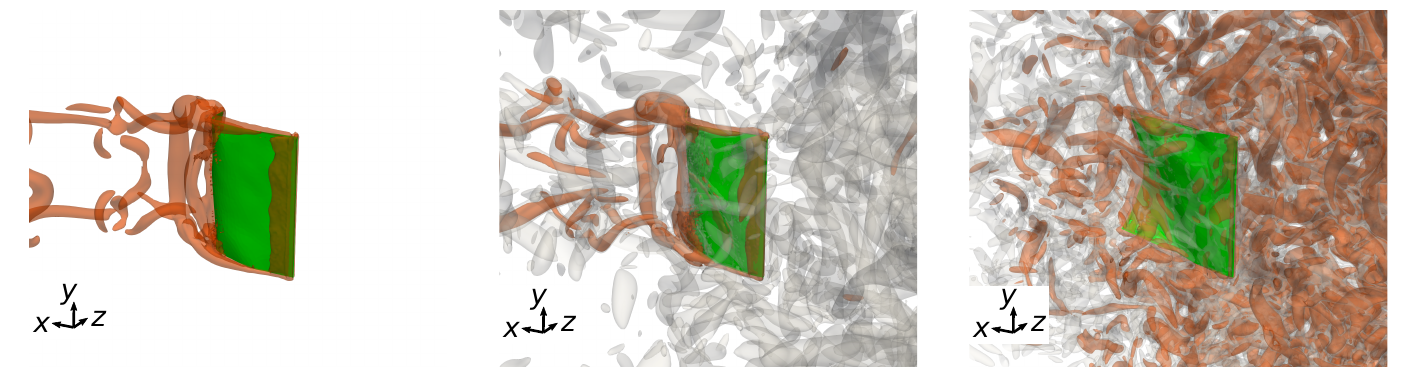}
\caption{
\added{Instantaneous vortical structures in the proximity of the flapping flag (green) at different turbulence intensities (left: $\mathit{Tu}=0$; centre: $\mathit{Tu}=20\%$; right: $\mathit{Tu}=40\%$). Isosurfaces detected using ${Q}$-criterion at two different levels are colored by white (lower intensity) and orange (higher intensity), respectively.}
}
\label{fig:snapshotFlag2}
\end{figure*}

\begin{figure*}
\centering
\includegraphics{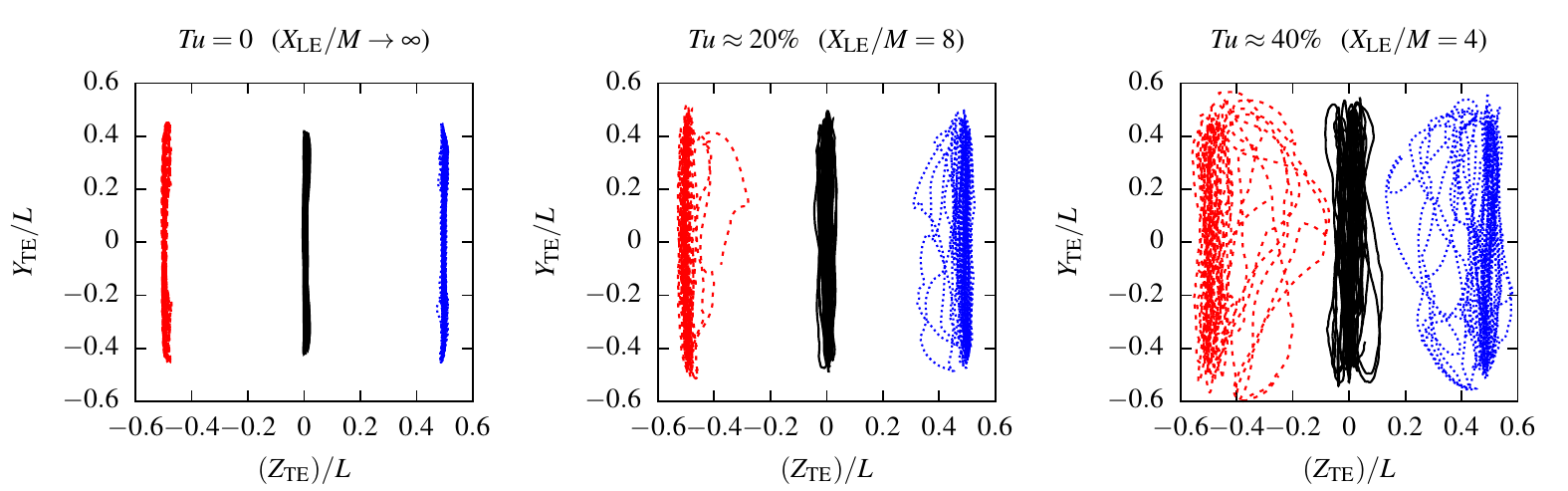} 
\caption{
Front-view trajectories of trailing edge middle (black) and side (red and blue) points at different turbulence intensities in the statistically steady state.
}
\label{fig:cfrTurb_TE-trajTrans}
\end{figure*}

\begin{figure*}
\centering
\includegraphics{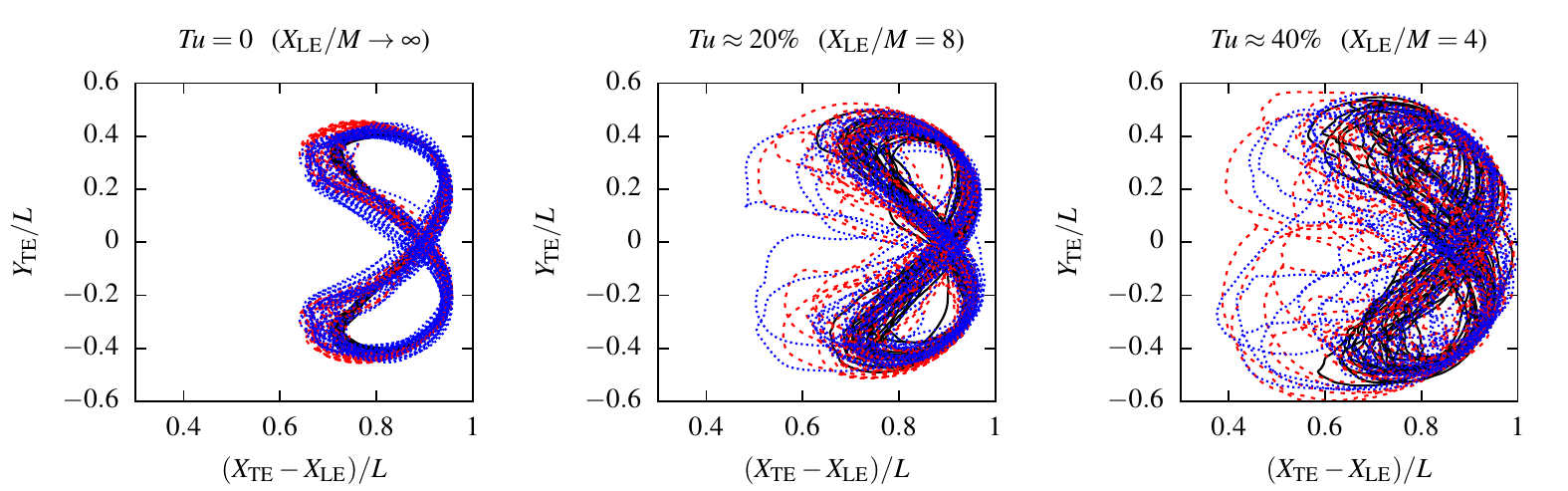}
\caption{
Side-view trajectories of trailing edge middle (black) and side (red and blue) points at different turbulence intensities in the statistically steady state.
}
\label{fig:cfrTurb_TE-traj}
\end{figure*}

\begin{figure*}
\centering
\includegraphics{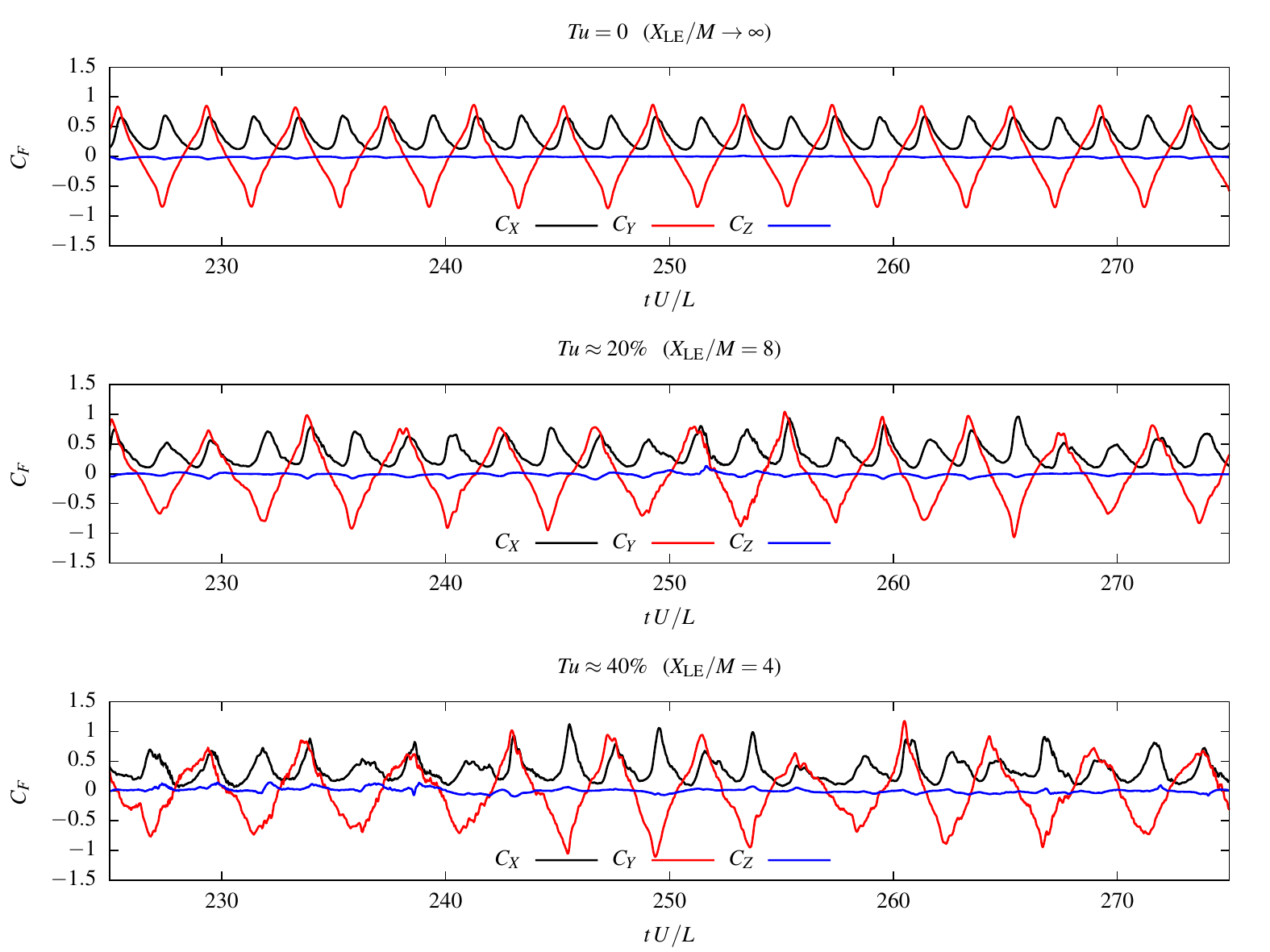}
\caption{
Time history of side (black), lift (blue) and drag (red) force coefficients at different turbulence intensities in the statistically steady state.
}
\label{fig:cfrTurb_cF}
\end{figure*}

To gain qualitative insight, Fig.~\ref{fig:snapshotFlag2} provides a visual comparison between the three cases focusing on the vortical flow structures detected using $Q$-criterion. In the laminar case ($\mathit{Tu}=0$), the characteristic wake pattern of the self-sustained flapping is obtained. In the intermediate case ($\mathit{Tu}=20\%$), such signature is still found, whereas it is eventually lost for the highest turbulence level ($\mathit{Tu}=40\%$). In the latter case, the flag deformation is more three-dimensional with a visible bending along the spanwise direction that is not present in the laminar case. 

The observation can be complemented by looking at the trajectories of the three points (i.e., middle and side points) from both the front and side view, shown in Fig.~\ref{fig:cfrTurb_TE-trajTrans} and Fig.~\ref{fig:cfrTurb_TE-traj}, respectively (we restrict now our attention to the statistically steady regime). An overall trend toward a more chaotic and three-dimensional motion is clearly observed. In particular, the characteristic shape of the limit-cycle oscillation (LCO) of the flapping flag in uniform flow is progressively lost as the turbulence level is increased. The same outcome can be obtained from the time histories of the aerodynamic coefficients reported in Fig.~\ref{fig:cfrTurb_cF}. Consistent\deleted{ly} with what observed for the TE motion, both the lift, side and drag coefficients become more irregular as the turbulence intensity increases.

\begin{figure*}
\centering
\includegraphics{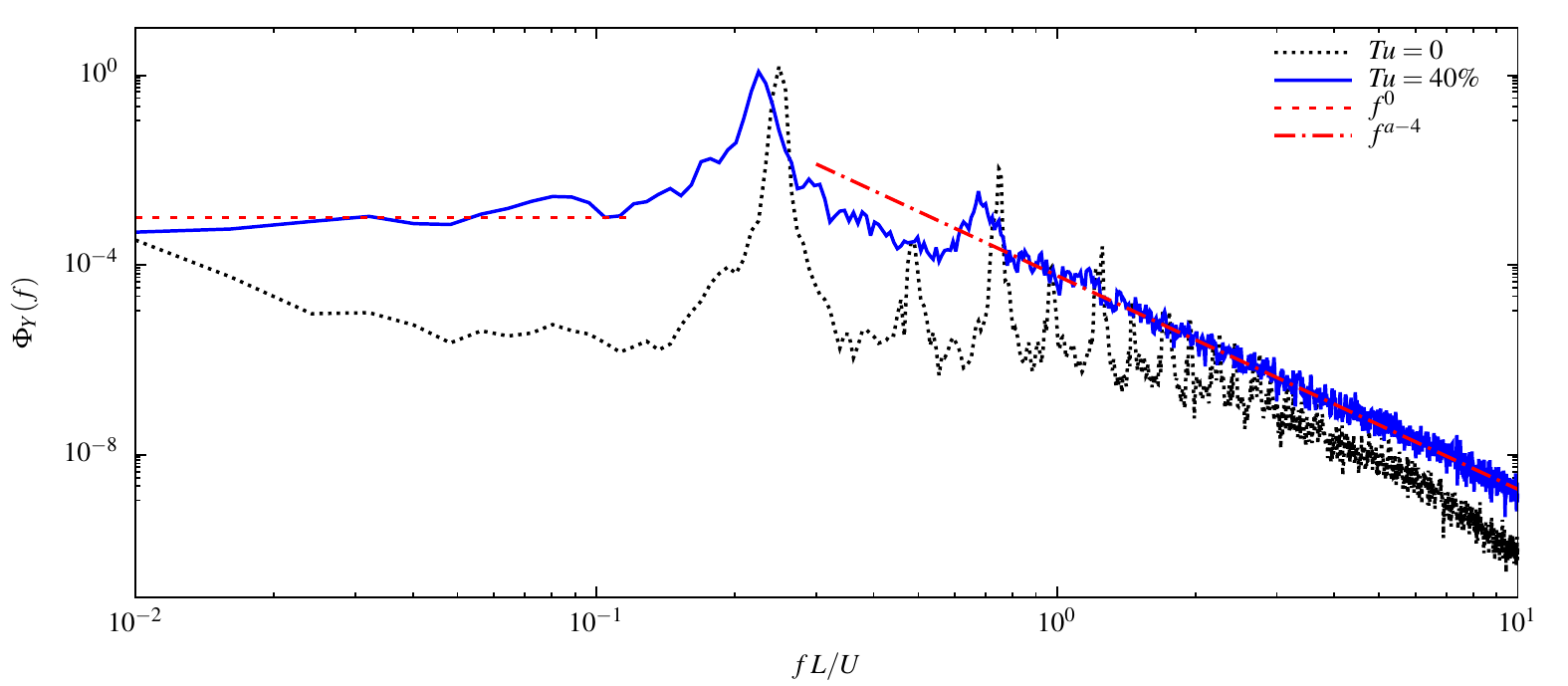}
\caption{
\added{Power spectral density of TE (mid point) transverse motion for the laminar ($\mathit{Tu}=0$) and the turbulent case at $\mathit{Tu}=40\%$. The latter is compared with the scaling laws predicted following the model by~\mbox{\citet{jin2016spectral}} (with the scaling exponent $a = -0.5$).}
}
\label{fig:cfrTurb_TE-spectra}
\end{figure*}

\begin{figure*}
\centering
\includegraphics{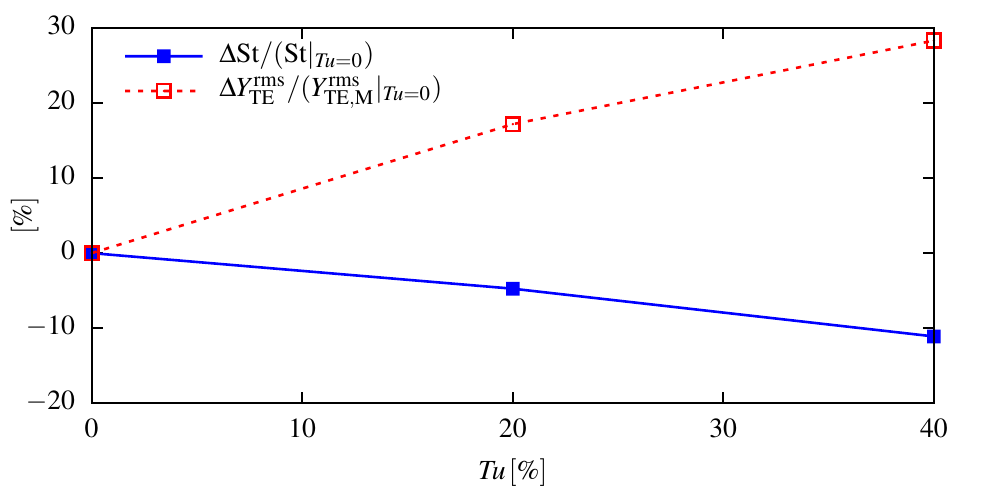}
\caption{
\added{Variation of Strouhal number and r.m.s. of TE (mid point) transverse oscillation with respect to the laminar case ($\mathit{Tu}=0$) as a function of the turbulence intensity.}
}
\label{fig:cfrTurb_St}
\end{figure*}

\added{
The effect of turbulence is reflected in the spectral content of the oscillation time history, shown in Fig.~\ref{fig:cfrTurb_TE-spectra}, where a much more broadband distribution can be observed for $\mathit{Tu}=40\%$ when compared to the laminar case ($\mathit{Tu}=0$), with a smoothing of the peak associated with the flapping frequency $f_\mathrm{flap}$. The latter, which can be expressed in nondimensional terms by the Strouhal number $\mathrm{St} = f_\mathrm{flap} L/U$, turns out to decrease with respect to the laminar case.
To qualitatively characterise the modification observed in the spectrum, we apply to our problem the dynamical model recently proposed by~\citet{jin2016spectral} for oscillating plates in active grid turbulence. By combining a forced-oscillator model and phenomenological arguments based on Kolmogorov theory, the following scalings for the oscillation spectrum can be argued~\cite{jin2016spectral}: (i) $\Phi_Y \propto f^0$ in the low-frequency region (i.e., for $f$ smaller than the natural frequency in the case of oscillating plate and such that the turbulence spectrum is approximately constant); (ii) $\Phi_Y \propto f^{-5/3-4}$ for the high-frequency region (i.e., for $f$ larger than the natural frequency and such that the turbulence spectrum is approximately within the inertial subrange). We underline that in our case the dominant frequency of the structural response is not the natural one but is determined by the nonlinear mutual interaction between the fluid and structural response. Moreover, the underlying assumptions of the model are not strictly met due to the limited Reynolds number (i.e., no clear inertial range can be observed) and distance of the flag from the grid (i.e., the flag is not properly within the decay region).
Therefore, in the high-frequency region we relax the assumption of inertial-range scaling and compare with a power-law $f^{a-4}$ (with $a=-0.5$) compatible with the numerical evidence, postponing to future investigations the more accurate characterization of the scaling exponent.
Notwithstanding, a reasonable agreement can be noticed in Fig.~\ref{fig:cfrTurb_TE-spectra} between the theoretical prediction and the numerical results for the turbulent case. On the other hand, for the laminar case a similar power-law subrange in the high-frequency region is absent.
These findings thus suggest that the fingerprint of turbulent fluctuations can be clearly detected from the resulting structural response.
Indeed, it has been recently shown that elastic dispersed objects can be exploited for measuring the two-point properties of turbulence~\cite{rosti2018flexible,rosti2020flowing,olivieri2021universal}.}

To conclude the analysis,  we look at how the Strouhal number and the flapping amplitude, i.e., the main quantities characterising the flapping motion, systematically vary with $\mathit{Tu}$. The relative change of these two observables is reported in Fig.~\ref{fig:cfrTurb_St}, where both quantities are normalised with the reference value obtained in the laminar case. As already noted, the Strouhal number decreases with the turbulence intensity. Conversely, the amplitude of the transverse motion, characterised in the chaotic case by the root-mean-square $Y_\mathrm{TE}^\mathrm{rms}$, is found to increase with $\mathit{Tu}$. A monotonic trend is observed for both $\mathrm{St}$ and $Y_\mathrm{TE}^\mathrm{rms}$, with the maximum variation of about $-10\%$ and $+30\%$ at the highest $\mathit{Tu}=40\%$, indicating that the effect of turbulence can also be relevant from a quantitative viewpoint.  

\section{Summary and perspectives}
\label{sec:conclusions}

In order to advance the current knowledge on fluid-structure interaction (mainly limited to the case of laminar flows), this work concerned the revisited study of the three-dimensional flag-in-the-wind problem in the framework of grid-induced turbulence. Grid turbulence is relevant from both the fundamental and applied viewpoint, but few numerical studies can be found in this regard due to the highly-demanding computational requirements imposed by the problem, in particular when considering the FSI simulation. To this aim, we performed fully-resolved direct numerical simulations exploiting state-of-the-art immersed boundary methods to properly describe the way in which turbulence is generated in the vicinity of the grid, as well as the coupled FSI dynamics of the flapping flag in both laminar and turbulent flow. The problem is tackled by taking advantage of the great potential of GPU-accelerated parallel computing and the computational framework here proposed is validated with respect to related experimental and numerical literature.

Specifically, we investigated the dynamics of a conventional flag (i.e., hinged at its leading edge) when immersed in the turbulence generated by regular passive grids at moderate Reynolds numbers. First, the turbulent flow was characterised in detail, including the evaluation of the power-law decay and the comparison with experimental measurements in similar conditions\added{\cite{djenidi2015power}}. Furthermore, the energy distribution was shown by means of a scale-by-scale analysis.  Then, we analysed the influence of the turbulent intensity on the flapping flag dynamics. We found that, although the overall trend towards a more chaotic motion is clearly noticed, the aeroelastic instability, and consequently the self-sustained flapping mechanism, still manifests also at the strongest turbulence level here considered. However, the effect of fluctuations impacting the body is such that the main features of the oscillations (including their amplitude and frequency) are remarkably affected. In particular, the flapping frequency (i.e., Strouhal number) is found to decrease whereas the amplitude increases with the turbulence intensity. Moreover, in agreement with recent experimental investigations~{\cite{jin2016spectral}, the fingerprint of turbulence can be detected and qualitatively described by looking at the spectral content of the altered oscillation. 

\added{
In addition, a point of particular interest, yet not addressed in the present study, is the comparison of the results from our fully-resolved approach with those alternatively obtained by introducing synthetic turbulence at the inflow. Although we can argue that the fully-resolved approach is more physically based since it involves well-defined geometrical and physical parameters, a more systematic and quantitative benchmark analysis represents a motivation for future work.}

In conclusion, our findings are relevant from a fundamental perspective, suggesting a general description for the enriched dynamical scenario of FSI in turbulence, as well as for several applications, such as flow control and energy harvesting in a more general configuration typical of real-world fluid flows.

\begin{acknowledgments}
The authors acknowledge fruitful discussions with Prof. Roberto Verzicco (University of Roma `Tor Vergata').
S.O. and M.E.R. acknowledge the computer time provided by the Scientific Computing section of Research Support Division at OIST. 
\added{A.M. acknowledges the financial support from the Compagnia di San Paolo, project MINIERA n. I34I20000380007, and from the project PoC -- BUYT MAIH, n. C36I20000140006.}
\end{acknowledgments}

\appendix

\section{Validation and convergence study}
\label{sec:convergence}

\begin{figure*}
\centering
\includegraphics{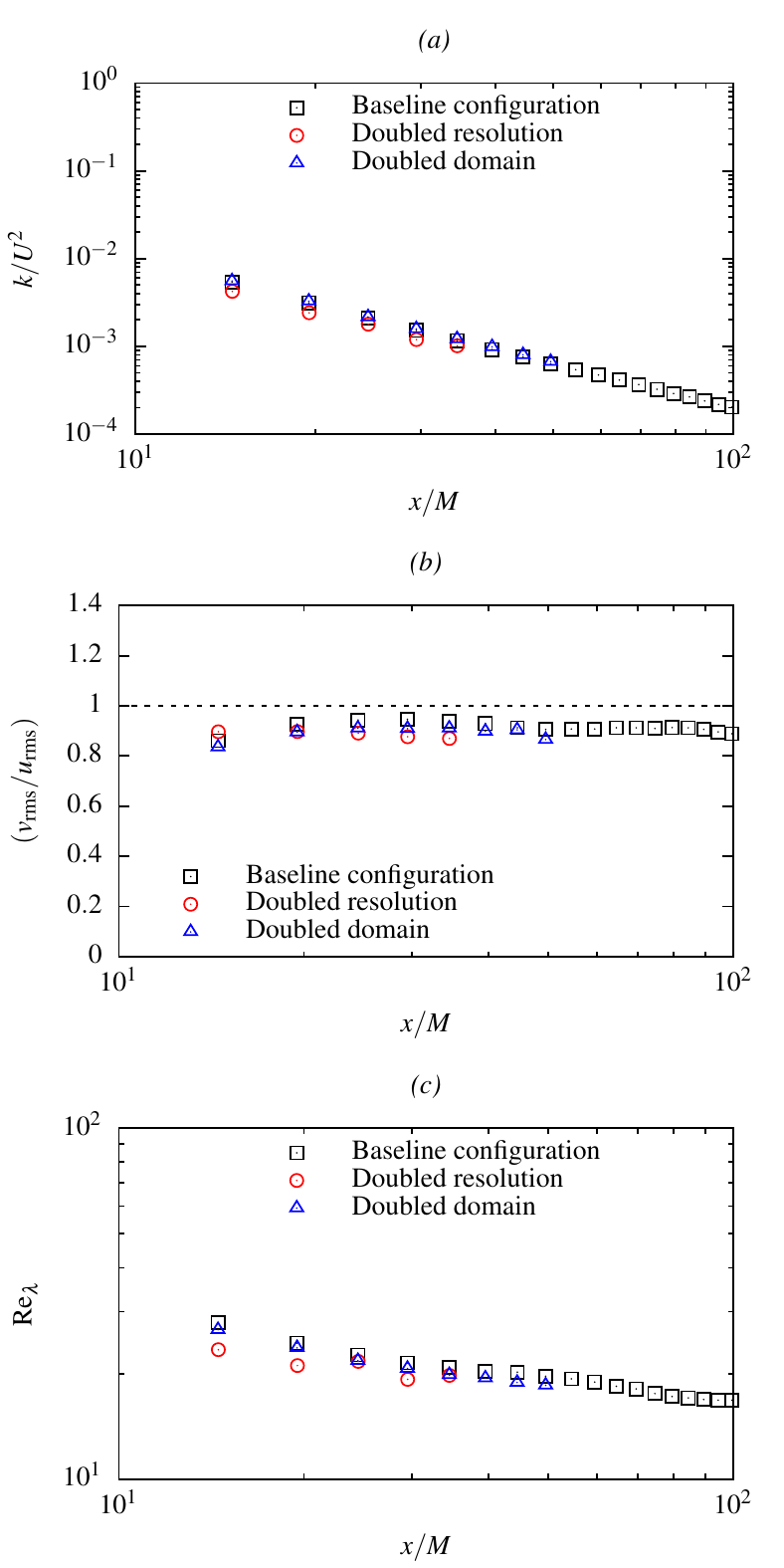}
\caption{
Convergence study on grid turbulence properties at $\mathrm{Re}_M=5000$ with respect to mesh resolution (red circles) and transverse domain extension (blue triangles) versus the baseline configuration adopted in the present work (black squares): \textit{(a)} turbulent kinetic energy; \textit{(b)} ratio between transverse and streamwise turbulent fluctuations; \textit{(c)} Reynolds number based on Taylor lengthscale. All quantities are reported as a function of the normalized streamwise coordinate.
}
\label{fig:convergence_1}
\label{fig:convergence_GT}
\end{figure*}

\subsection{Grid-induced turbulence}
Concerning the numerical characterisation of grid-induced turbulence (Sec.~\ref{sec:flow}), the sensitivity of our results has been checked with respect to both spatial and temporal resolution, as well as to the transverse domain extension. Here, we focus on the former and the latter, since the influence of decreasing the simulation timestep was found to be negligible (not shown). A convergence analysis has been performed considering the case $\mathrm{Re}_M=5000$. For the spatial resolution, we have performed a simulation where the mesh spacing $\Delta x$ is halved, i.e., doubling the resolution of the baseline configuration. For the transverse domain extension, we doubled both $L_y$ and $L_z$ while maintaining the same mesh spacing. To retain a feasible computational time while increasing the number of mesh points, we have decreased the streamwise domain extension to $L_x/M=40$ and 60, respectively. In order to have a faster statistical convergence, we take advantage of the increased number of points and perform a spatial average for the quantities of interest over the transverse slices.

Figure~\ref{fig:convergence_1} shows the results from these tests compared with those obtained using the baseline configuration adopted in the present work, in terms of the same quantities presented in Sec.~\ref{sec:flow}, i.e., turbulent kinetic energy (Fig.~\ref{fig:convergence_1}\textit{(a)}), large-scale anisotropy (Fig.~\ref{fig:convergence_1}\textit{(b)}) and turbulent Reynolds number (Fig.~\ref{fig:convergence_1}\textit{(c)}). It can be noticed that the numerical results are overall robust. In particular, the influence of the domain extension appears to be minimal. On the other hand, the variation observed when refining the mesh resolution is more evident, although still small. Indeed, the latter can be linked to the aforementioned dependency of the generated turbulence on the small-scale details associated with boundary-layer effects and flow separation in the proximity of the grid.
\added{In fact, by comparing the smallest values of $\eta$ (shown in Fig.~\ref{fig:charLen}) with the baseline mesh spacing, it can be argued that the simulation is slightly under-resolved close to the grid.
 Nevertheless, we remark that the results obtained with the baseline configuration are in good agreement with experimental measurements using a similar setting (see Fig.~\ref{fig:decay2}), therefore confirming that the chosen resolution is adequate to describe the essential features (especially when focusing on the large- and meso-scale features) of the turbulent flow in the region of interest.}

\begin{figure*}
\centering
\includegraphics{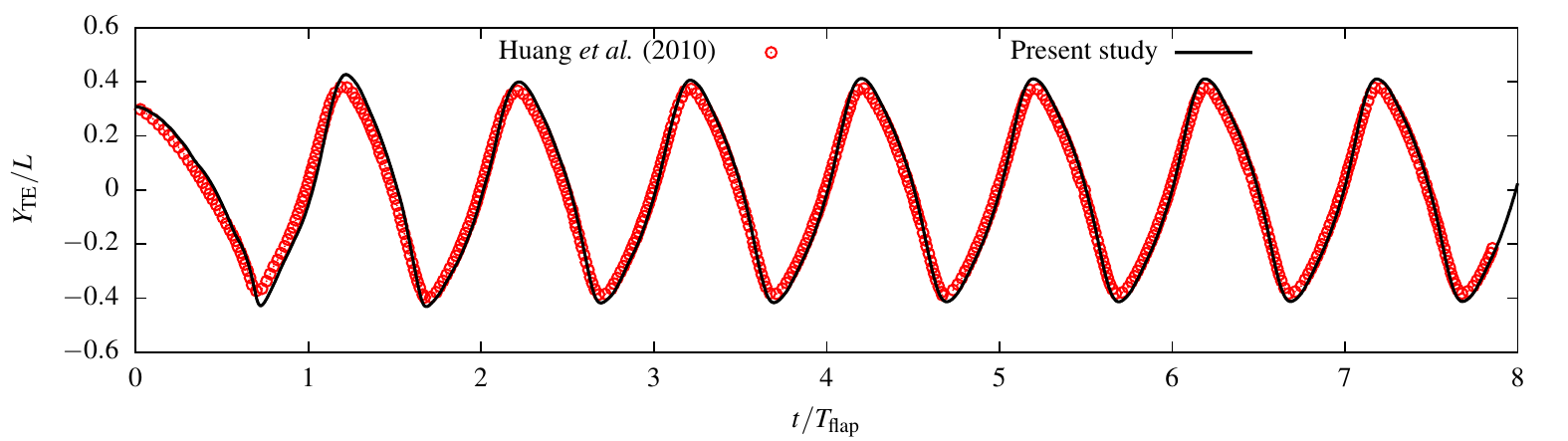}
\caption{
Time history of the trailing edge mid point for flapping flag in laminar flow at $\mathrm{Re_L} = 200$. Comparison between the present results (black line) and those by~\citet{huang2010three} (red symbols).
}
\label{fig:flagVal}
\end{figure*}

\subsection{Flapping flag in laminar flow}
\label{sec:flagVal}

To check the accuracy of the FSI simulation, we validate our numerical technique by considering the coupled motion of the flapping flag in a uniform flow at $\mathrm{Re}=200$. This configuration has been originally investigated by~\citet{huang2010three} and later also by Refs.~\cite{tian2014fluid,lee2015discrete,de2016moving}. Focusing on the time history of the transverse oscillation (we consider only the TE mid point since the motion is essentially two-dimensional), Fig.~\ref{fig:flagVal} shows a comparison between our results and those by~\citet{huang2010three}. Here, time is normalised using the corresponding flapping period (differing only about 2\%). A good agreement can be observed both in the shape and amplitude of the oscillation indicating that the numerical procedure is accurately reproducing the reference solution. 

\section*{Data Availability}
The data that support the findings of this study are available from the corresponding author upon reasonable request.


\providecommand{\noopsort}[1]{}\providecommand{\singleletter}[1]{#1}%

\end{document}